\documentclass[aps,prd,reprint,nofootinbib,superscriptaddress,showpacs]{revtex4-1}

\usepackage{amsmath,amssymb,amsfonts}
\usepackage[utf8]{inputenc}
\usepackage{graphicx}
\usepackage{url}
\usepackage{hyperref}
\usepackage{bm,bbm}
\usepackage{booktabs}

\newcommand{\nwse}[3]{\ensuremath{#1^{#2}_{\phantom{#2} #3}}}

\newcommand{\ucsc}{Departamento de Matemática y Física Aplicadas, Universidad Católica de la Santísima Concepción, Alonso de Ribera 2850, 4090541 Concepción, Chile}
\newcommand{\cbpf}{Centro Brasileiro de Pesquisas F\'{\i}sicas, Rua~Dr.~Xavier Sigaud, 150, Urca, 22290-180 Rio de Janeiro - RJ, Brazil}
\newcommand{\uval}{Departamento de Física Teórica, Atómica y Óptica, Universidad de Valladolid, Calle del Doctor Mergelina s/n, 47011-Valladolid, Spain}

\begin{document}

\title{Effectively four-dimensional spacetimes emerging from\\$d=5$ Einstein--Gauss--Bonnet Gravity}

\author{Fernando Izaurieta}
\email{fizaurie@ucsc.cl}
\affiliation{\ucsc}
\affiliation{\uval}

\author{Eduardo Rodr\'{\i}guez}
\email{edurodriguez@ucsc.cl}
\affiliation{\ucsc}
\affiliation{\cbpf}

\date{\today}

\begin{abstract}
Einstein--Gauss--Bonnet gravity in five-dimensional spacetime provides an excellent example of a theory that,
while including higher-order curvature corrections to General Relativity, still shares many of its features, such as second-order field equations for the metric.
We focus on the largely unexplored case where the coupling constants of the theory are such that no constant-curvature solution is allowed, leaving open the question of what the vacuum state should then be.
We find that even a slight deviation from the anti-de~Sitter Chern--Simons theory, where the vacuum state is five-dimensional AdS spacetime, leads to a complete symmetry breakdown, with the fifth dimension either being compactified into a small circle or shrinking away exponentially with time.
A complete family of solutions, including duality relations among them, is uncovered and shown to be unique within a certain class.
This dynamical dimensional reduction scenario seems particularly attractive as a means for higher-dimensional theories to make contact with our four-dimensional world.
\end{abstract}


\maketitle

\section{Introduction}
\label{sec:intro}

S.~Carroll~\cite{Car07} once likened the cosmological constant to Rasputin --- difficult to kill off.
From a purely theoretical point of view, the cosmological constant has a singular appeal; it is the only diffeomorphism-invariant term one can add to the Ricci scalar curvature in the four-dimensional Einstein--Hilbert (EH) action that does not spoil the second-order nature of the field equations for the metric.
By the same token, omitting the cosmological constant from the action requires an explanation --- why set $\Lambda = 0$?

Gravity theories in dimensions higher than four that go beyond the EH action have been known at least since the work of D.~Lovelock~\cite{Lov71}.
Their rationale is simple.
Any diffeomorphism-invariant term that leads to second-order field equations for the metric is allowed in the action.
In $d$ dimensions, the most general Lovelock Lagrangian can be written as a linear combination of
\begin{equation}
 L_{p}^{\left( d \right)} = \frac{1}{2^{p}}
 \delta^{\nu_{1} \cdots \nu_{2p}}_{\mu_{1} \cdots \mu_{2p}}
 \nwse{R}{\mu_{1} \mu_{2}}{\nu_{1} \nu_{2}} \cdots
 \nwse{R}{\mu_{2p-1} \mu_{2p}}{\nu_{2p-1} \nu_{2p}},
\end{equation}
where $\nwse{R}{\mu \nu}{\rho \sigma}$ is the Riemann tensor
and $p = 0, 1, \ldots, n = \left\lfloor d / 2 \right\rfloor$.
The first two terms are the cosmological constant ($p=0$) and EH ($p=1$) terms.
By disentangling the Kronecker delta we can check that the $p=2$ term amounts to the exact combination of curvature-squared contractions that define the Gauss--Bonnet (GB) density,
\begin{equation}
 L_{2}^{\left( d \right)} =
 \nwse{R}{\mu \nu}{\rho \sigma} \nwse{R}{\rho \sigma}{\mu \nu} - 4 \nwse{R}{\mu}{\nu} \nwse{R}{\nu}{\mu} + R^{2}.
 \label{eq:GB}
\end{equation}
Adding this term to the EH action in $d=4$ does not change the field equations,
because it can be written as a total derivative that only contributes a boundary term to the action~\cite{Lan38}.%
\footnote{Noether charges are sensitive to boundary terms in the action, however, and the inclusion of $L_{2}^{\left( 4 \right)}$ has been shown to be important in this context~\cite{Aro99a,Aro99b}.}
Things are different when $d \geq 5$.
The inclusion of the higher-order terms in dimensions greater than four produces a stark departure from General Relativity (GR),
even though the principles behind their inclusion are the same as those for the EH action.
As with the cosmological constant, it would seem that omitting the terms with $p \geq 2$ from the action requires an explanation.
That there may be none available is highlighted by the fact that these higher-order corrections to GR are actually predicted by String Theory~\cite{Zwi85,Bou85,Zum86,Wil86}.

The lowest dimension in which the Lovelock terms make a difference,
and hence the simplest scenario where we can test their consequences, is $d=5$.
Five-dimensional spacetime is also interesting from a rather different point of view, namely, the AdS/CFT correspondence~\cite{Aha99}, with the prime example being the equivalence of $\mathcal{N}=4$, $U \left( N \right)$ Yang--Mills theory and ten-dimensional superstring theory on $AdS_{5} \times S_{5}$.
Einstein--Gauss--Bonnet (EGB) gravity, which includes up to the quadratic Lovelock term, has been extensively studied in five and higher dimensions%
\footnote{Other theories that also carry the ``Gauss--Bonnet'' label include $f \left( G \right)$ theories, where the EH action is supplemented with some nontrivial function of the GB term, and ``dilatonic'' GB theories, where the departure from GR comes from a scalar field that couples non-minimally to the GB term (see, e.g., Ref.~\cite{Noj06}). We shall restrict ourselves to EGB gravity as summarized in section~\ref{sec:EGB}.}
(among the vast literature, see, e.g., Refs.~\cite{Mad85,Mul85,Whe86a,Whe86b,Cve01,Cai01,Cho01,Cha02,Can07a,Can07b,Dot10,Mae11b,Gol12,Bri12}).

\begin{figure}[thpb]
 \centering
 \includegraphics[width=.6\columnwidth]{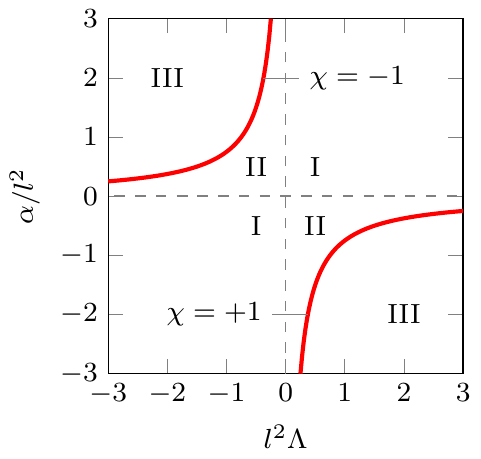}
 \caption{The two branches of the hyperbola $\alpha \Lambda = -3/4$ define different regions of parameter space for EGB gravity in $d=5$ (the picture is qualitatively the same in higher dimensions). Regions~I and~II admit two distinct constant-curvature solutions, while region~III admits none. On the boundary between regions~II and~III lies the CS case, where a single constant-curvature vacuum exists. Our parameterization for the EGB coupling constants, which replaces
$\left( \alpha, \Lambda \right)$ with $\left( l, \chi \right)$
[cf.~eqs.~(\ref{eq:Lambda})--(\ref{eq:alpha})], is valid in regions~II and~III.}
 \label{fig:phase}
\end{figure}

It has long been recognized that the EGB theory possesses different regimes depending on the value of the product $\alpha \Lambda$, where $\alpha$ is the GB coupling constant.
These different regions of parameter space are shown graphically in figure~\ref{fig:phase}.
In regions~I and~II there exist two distinct constant-curvature solutions, each of which is a candidate for the vacuum~\cite{Bou85,Des89,Des02a,Des02b,Des07,Cha08}.
On the boundary between regions~II and~III, $\alpha$ and $\Lambda$ conspire to let the action acquire an extra symmetry beyond local Lorentz invariance.
This is the Chern--Simons (CS) case~\cite{Cha89}, and the extra symmetry may be de~Sitter (dS), anti-de~Sitter (AdS) or Poincaré invariance.%
\footnote{In this latter case only the GB term is present in the action.}
One of the beauties of the CS case is that the theory can be seen to have a single vacuum state --- a constant-curvature spacetime.
In region~III, on the other hand, no constant-curvature solution is allowed.

Perhaps unsurprisingly, this last case, where no constant curvature vacuum may exist, has received little attention so far.
The absence of a maximally symmetric, constant-curvature solution may, on the other hand, signal the existence of a less symmetric vacuum state.
For instance, spacetime may be forced to acquire a ``warped product'' structure, factorizing into, e.g., $4+1$ or $3+2$ spaces.
A less symmetric vacuum state may then turn out to look a lot like our own usual four-dimensional spacetime, with the fifth dimension being dynamically compactified away or otherwise rendered unobservable.

Our motivation in this work has precisely been to look for solutions of EGB gravity that display a warped product structure that somehow permits an effectively four-dimensional spacetime to emerge as the non-maximally symmetric vacuum state of this five-dimensional theory.

In section~\ref{sec:EGB} we review the EGB theory in greater detail and set the stage for our search.
Our main results are presented in summarized form in section~\ref{sec:sols}.
We find a class of very simple exact solutions and in section~\ref{sec:examples} we focus our attention on two of them.
The first one realizes dynamically the Kaluza--Klein idea of a circular fifth dimension,
with the flow of time changing as we move along the fifth dimension.
In the second solution the fifth dimension is non-compact, but it shrinks away exponentially into nothingness as time passes, leaving behind an effectively four-dimensional spacetime.
Ordinary three-dimensional space for both solutions has constant negative curvature.
These solutions show, in as simple a setting as possible, how an effectively four-dimensional spacetime may dynamically emerge from a higher-dimensional theory.
Note that four-dimensional spacetime emerges as a natural feature of the theory, and not as the result of some ad-hoc compactification.

In section~\ref{sec:prod} we briefly study our solutions from a geometrical point of view and show that their maximal extensions correspond to the product of two constant-curvature manifolds of dimension two and three, respectively.
We then go on to show that no other possible product of constant-curvature manifolds is allowed as a vacuum solution of EGB gravity in region~III.
We state our conclusions and further discuss our results in section~\ref{sec:final}, where we also give an outlook for future work.

\section{The Vacuum in Einstein--Gauss--Bonnet Gravity}
\label{sec:EGB}

For our purposes it will prove useful to write the Lagrangian for five-dimensional EGB gravity in the language of differential forms as%
\footnote{The usual definitions of first-order formulations for gravity apply here:
$e^{a}$ and $\omega^{ab}$ stand for the one-forms vielbein and spin connection,
with the Lorentz curvature and torsion being defined as
$R^{ab} = \mathrm{d} \omega^{ab} + \nwse{\omega}{a}{c} \omega^{cb}$,
$T^{a} = \mathrm{d} e^{a} + \nwse{\omega}{a}{b} e^{b}$.
Wedge product between differential forms is understood throughout.}
\begin{equation}
L =
\frac{\kappa}{l}
\epsilon_{abcde}
\left(
  R^{ab} R^{cd}
  - \frac{2 \chi}{3 l^{2}} R^{ab} e^{c} e^{d}
  + \frac{1}{5 l^{4}} e^{a} e^{b} e^{c} e^{d}
\right)
e^{e}.
\label{eq:LEGB}
\end{equation}
Our choice of parameterization for the coupling constants deserves some explanation.
The only dimensionful constant is $l$, which matches every appearance of the vielbein and has dimensions of length.
Clearly $l$ could be eliminated by absorbing it in the definition of the vielbein,
but then the spacetime metric $g_{\mu \nu}$ would no longer be related to $e^{a}$ through the familiar relation
$g_{\mu \nu} = \eta_{ab} \nwse{e}{a}{\mu} \nwse{e}{b}{\nu}$.
The constants $\kappa$ and $\chi$ are dimensionless;%
\footnote{In natural units, with $\hbar = c = 1$.}
the first one is related to Newton's constant while the latter measures the relative strength of the EH term in eq.~(\ref{eq:LEGB}).
A larger absolute value of $\chi$ is thus associated with greater similarity with GR.
The cosmological constant $\Lambda$ and the GB coupling constant $\alpha$ can be recovered from $l$ and $\chi$ through the relations
\begin{align}
\Lambda & = \frac{3}{\chi l^{2}},  \label{eq:Lambda} \\
\alpha & = - \frac{l^{2}}{4 \chi}. \label{eq:alpha}
\end{align}
As inferred from eqs.~(\ref{eq:Lambda})--(\ref{eq:alpha}), our parameterization is strictly speaking only valid when $\Lambda$ and $\alpha$ have \emph{different} signs (regions~II and~III in figure~\ref{fig:phase}).
Since our interest here lies in the case when no constant-curvature vacuum exists, this will prove sufficient --- it can be easily shown that, when $\Lambda$ and $\alpha$ have the \emph{same} sign (region~I in figure~\ref{fig:phase}), then the theory always admits two distinct constant-curvature solutions.

The field equations obtained by independently varying the Lagrangian~(\ref{eq:LEGB}) with respect to the vielbein and the spin connection read
\begin{align}
\epsilon_{abcde}
\left(
  R^{ab} R^{cd}
  - \frac{2 \chi}{l^{2}} R^{ab} e^{c} e^{d}
  + \frac{1}{l^{4}} e^{a} e^{b} e^{c} e^{d}
\right)
& = 0,
\label{eq:e} \\
\epsilon_{abcde}
\left(
  R^{cd}
  - \frac{\chi}{l^{2}} e^{c} e^{d}
\right)
T^{e}
& = 0.
\label{eq:w}
\end{align}
While eq.~(\ref{eq:w}) can be solved by simply demanding $T^{a} = 0$, and we shall do so in everything that follows, it seems worthwhile to recall that there exist solutions to the EGB equations that feature nontrivial torsion~\cite{Can07a}.
Eq.~(\ref{eq:e}) is the EGB generalization of the Einstein equation in vacuum.
The presence of the GB term in the action implies that it can be factorized as
\begin{equation}
\epsilon_{abcde}
\left(
  R^{ab} - \beta_{+} e^{a} e^{b}
\right)
\left(
  R^{cd} - \beta_{-} e^{c} e^{d}
\right)
= 0,
\end{equation}
where the roots $\beta_{\pm}$ are given by
\begin{equation}
\beta_{\pm}
= \frac{\chi}{l^{2}}
\left(
  1 \pm
  \sqrt{1 - \frac{1}{\chi^{2}}}
\right).
\end{equation}
This simple calculation shows that the behavior of the EGB theory regarding constant-curvature solutions,
$R^{ab} = \beta_{\pm} e^{a} e^{b}$,
can be summarized as follows.
When $\chi^{2} = 1$ there exists a single constant-curvature vacuum; dS spacetime for $\chi = +1$ and AdS spacetime for $\chi = -1$.
This case is also special in the sense that the EGB Lagrangian coincides with the CS form for the dS algebra (when $\chi = +1$) or the AdS algebra (when $\chi = -1$).
When $\chi^{2} > 1$ (regions~I and~II in figure~\ref{fig:phase}) there exist two distinct constant-curvature solutions, each of which is a candidate for the vacuum~\cite{Bou85,Des89,Des02a,Des02b,Des07,Cha08}.
A last possibility occurs when $\chi^{2} < 1$ (region~III in figure~\ref{fig:phase}) and the roots $\beta_{\pm}$ are complex, with $\beta_{\mp} = \beta_{\pm}^{\ast}$.
In this case there are no constant-curvature solutions, so the vacuum state is bound to have less than maximal symmetry.
We would like to emphasize that, while $\beta_{\pm}$ are complex in the latter case, the Lagrangian stays always real.

As mentioned in the Introduction, we shall be particularly concerned with the case where no constant-curvature solutions exist, i.e., with the case where $\chi^{2} < 1$ (region~III in figure~\ref{fig:phase}).

\section{Ansatz and Solutions}
\label{sec:sols}

We begin our search for solutions by postulating the metric ansatz
\begin{equation}
ds^{2} =
- f^{2} \left( w \right) dt^{2}
+ g^{2} \left( w \right) d\Sigma^{2}
+ p^{2} \left( t \right) q^{2} \left( x,y,z \right) dw^{2},
\label{eq:ansatz}
\end{equation}
where $d\Sigma$ stands for the line element of a constant-curvature 3-manifold $\Sigma$,
\begin{equation}
d\Sigma^{2} =
\left[ 1 + \frac{K}{4} \left( x^{2} + y^{2} + z^{2} \right) \right]^{-2}
\left( dx^{2} + dy^{2} + dz^{2} \right).
\label{eq:Sigma}
\end{equation}
The ansatz in eqs.~(\ref{eq:ansatz})--(\ref{eq:Sigma}) includes four unknown functions,
$f \left( w \right)$,
$g \left( w \right)$,
$p \left( t \right)$,
$q \left( x,y,z \right)$, and one parameter, $K$.
The warped product structure is apparent in the fact that functions of $w$ multiply $dt$ and the line element of $\Sigma$,
while functions of $t$ and $\left( x,y,z \right)$ multiply $dw$.

To plug this ansatz into the field equations~(\ref{eq:e}) we need to first turn it from a metric ansatz into an ansatz for the vielbein and the spin connection.
The vielbein part can be chosen as
\begin{align}
 e^{0} & = f \left( w \right) \mathrm{d} t, \label{eq:e0} \\
 e^{i} & = g \left( w \right) \tilde{e}^{i}, \\
 e^{4} & = p \left( t \right) q \left( x,y,z \right) \mathrm{d} w, \label{eq:e4}
\end{align}
where $\tilde{e}^{i}$ ($i=1,2,3$) stands for the intrinsic vielbein of the three-manifold $\Sigma$,
\begin{equation}
 \tilde{e}^{i} =
 \left[ 1 + \frac{K}{4} \left( x^{2} + y^{2} + z^{2} \right) \right]^{-1} \mathrm{d} x^{i}.
\end{equation}
The ansatz for the spin connection, on the other hand, can be found by solving $T^{a}=0$ for $\omega^{ab}$ and plugging eqs.~(\ref{eq:e0})--(\ref{eq:e4}) into the result.
We shall spare the reader the (not particularly illuminating) details.

The field equations [cf.~eq.~(\ref{eq:e})] for the EGB theory are notoriously complicated,
amounting to a highly nonlinear system of 25 coupled partial differential equations.
It is remarkable that the ansatz~(\ref{eq:ansatz}) allows for an exact solution to be found.

Our results imply that we must have
\begin{align}
g \left( w \right) & = 1, \\
q \left( x,y,z \right) & = 1, \\
K & = \frac{1}{\chi l^{2}}.
\end{align}
Surprisingly, the field equations do not allow for a nontrivial warping factor for the three-manifold $\Sigma$,
while the warping factor for the fifth dimension is only allowed to depend on time.
This means that the metric takes on the simplified form
\begin{equation}
ds^{2} =
- f^{2} \left( w \right) dt^{2}
+ d\Sigma^{2}
+ p^{2} \left( t \right) dw^{2}.
\label{eq:simpansatz}
\end{equation}
All possible alternatives for the $f$ and $p$ functions allowed by the EGB field equations are summarized in table~\ref{tab:alt}.

\begin{table}[thpb]
 \caption{\label{tab:alt}Summary of solutions for the EGB theory with $\chi^{2} < 1$. Each spacetime is characterized by the functions $f$ and $p$ plus the sign of the curvature of $\Sigma$. Here ``circ.'' refers to a linear combination of sine and cosine, while ``hyp.'' refers to a linear combination of hyperbolic sine and hyperbolic cosine. The parameters $R$ and $\tau$ set length and time scales, respectively, with $\xi \equiv \left( 1/2 \right) \left( \chi - 1/\chi \right)$.}
 \begin{ruledtabular}
 \begin{tabular}{cccccc}
  Class & $f \left( w \right)$ & $p \left( t \right)$ & $\Sigma$ & Range & $R,\tau$ \\
  \hline
  PH$-$ & 1 & hyp.  & $K<0$ & $-1 < \chi < 0$ & $l \sqrt{\xi}$ \\
  FC$-$ & circ. & 1 & $K<0$ & $-1 < \chi < 0$ & $l \sqrt{\xi}$ \\
  PC$+$ & 1 & circ. & $K>0$ & $0 < \chi < 1$  & $l \sqrt{-\xi}$ \\
  FH$+$ & hyp. & 1  & $K>0$ & $0 < \chi < 1$  & $l \sqrt{-\xi}$ \\
 \end{tabular}
 \end{ruledtabular}
\end{table}

We have adopted a very simple-minded naming scheme for these solutions.
A ``PH$-$'' solution, for instance, has hyperbolic $p$ (and trivial $f$), with negative curvature for $\Sigma$.
A ``PC$+$'' solution, on the other hand, has circular $p$ (and trivial $f$), with positive curvature for $\Sigma$.
As show in table~\ref{tab:alt}, length and time scales for all these solutions are derived by means of a new parameter,
\begin{equation}
 \xi = \frac{1}{2} \left( \chi - \frac{1}{\chi} \right).
 \label{eq:xi}
\end{equation}

It is clear from eq.~(\ref{eq:xi}) that, as $\chi \rightarrow 1/\chi$, we have $\xi \rightarrow -\xi$.
This transformation on $\chi$ leads us outside the range of interest that we had set for $\chi$ at the beginning,
namely, from $\chi^{2} < 1$ to $\chi^{2} > 1$.
The solutions carry over.
As shown in table~\ref{tab:alt2}, four more classes of solutions similar to the ones already discussed also exist in the theory with $\chi^{2} > 1$, and, perhaps not so surprisingly, also for GR with either a positive or a negative cosmological constant.
Existence, however, does not necessarily imply relevance.
While the solutions in table~\ref{tab:alt} for the EGB theory with $\chi^{2} < 1$ are strong candidates for the vacuum state of the theory, both GR and the EGB theory with $\chi^{2} > 1$ have their own vacua --- five-manifolds of constant curvature.

\begin{table}[thpb]
 \caption{\label{tab:alt2}Summary of solutions for the EGB theory with $\chi^{2} > 1$ and GR with either a positive or a negative cosmological constant $\Lambda$. See also table~\ref{tab:alt} and figure~\ref{fig:sols}.}
 \begin{ruledtabular}
 \begin{tabular}{cccccccc}
  Class & $f \left( w \right)$ & $p \left( t \right)$ & $\Sigma$ & Theory & Range & $R,\tau$ & $K$ \\
  \hline
  PH$+$ & 1     & hyp.  & $K>0$ & EGB & $\chi > 1$    & $l \sqrt{\xi}$               & $\frac{1}{\chi l^2}$ \\
        &       &       &       & GR  & $\Lambda > 0$ & $\sqrt{\frac{3}{2\Lambda}}$  & $\frac{\Lambda}{3}$ \\
  \hline
  FC$+$ & circ. & 1     & $K>0$ & EGB & $\chi > 1$    & $l \sqrt{\xi}$               & $\frac{1}{\chi l^2}$ \\
        &       &       &       & GR  & $\Lambda > 0$ & $\sqrt{\frac{3}{2\Lambda}}$  & $\frac{\Lambda}{3}$ \\
  \hline
  PC$-$ & 1     & circ. & $K<0$ & EGB & $\chi < -1$   & $l \sqrt{-\xi}$              & $\frac{1}{\chi l^2}$ \\
        &       &       &       & GR  & $\Lambda < 0$ & $\sqrt{-\frac{3}{2\Lambda}}$ & $\frac{\Lambda}{3}$ \\
  \hline
  FH$-$ & hyp.  & 1     & $K<0$ & EGB & $\chi < -1$   & $l \sqrt{-\xi}$              & $\frac{1}{\chi l^2}$ \\
        &       &       &       & GR  & $\Lambda < 0$ & $\sqrt{-\frac{3}{2\Lambda}}$ & $\frac{\Lambda}{3}$ \\
 \end{tabular}
 \end{ruledtabular}
\end{table}

It is perhaps interesting to note that, as $\chi \rightarrow \infty$ and $l \rightarrow 0$,
while keeping $\chi l^{2} = 3 / \Lambda$ finite,
the EGB solutions with $\chi > 1$ exactly match those of GR with $\Lambda > 0$.
An analogous phenomenon occurs for the EGB theory with $\chi < -1$ and GR with $\Lambda < 0$.

\begin{figure}[thpb]
 \centering
 \includegraphics[width=.9\columnwidth]{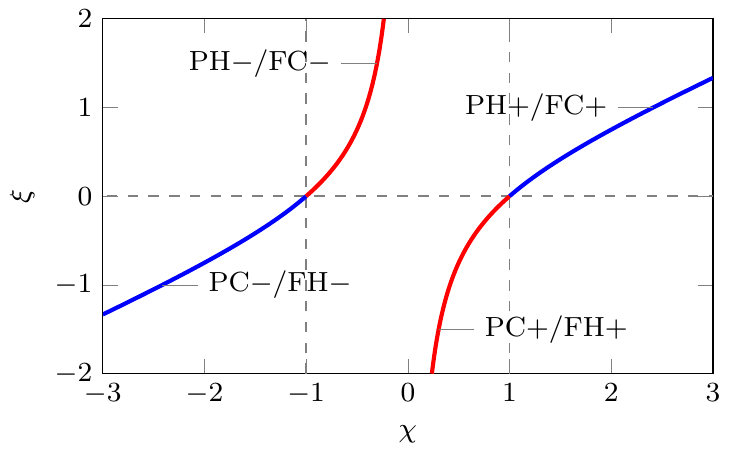}
 \caption{Summary of solutions as parameterized by $\chi$ and $\xi$, with $\xi \equiv \left( 1/2 \right) \left( \chi - 1/\chi \right)$. Solutions for the EGB theory with $\chi^{2} > 1$ are also solutions of GR, whereas those with $\chi^{2} < 1$ are not. See also table~\ref{tab:alt} and table~\ref{tab:alt2}.}
 \label{fig:sols}
\end{figure}

The full picture of solutions, as parameterized by $\chi$ and $\xi$, is shown in figure~\ref{fig:sols}.

\section{Five going on Four}
\label{sec:examples}

\subsection{Circular Fifth Dimension}
\label{sec:circ}

An FC$-$ solution (see table~\ref{tab:alt} and figure~\ref{fig:sols}) can always be written as
\begin{equation}
ds^{2} = - \cos^{2} \left( \frac{w}{R} \right) dt^{2} + d\Sigma^{2} + dw^{2},
\label{eq:FW}
\end{equation}
where $R = l \sqrt{\xi}$ [cf.~eq.~(\ref{eq:xi})] and $\Sigma$ has constant negative curvature.

The spacetime described by the line element~(\ref{eq:FW}) is one where the fifth dimension has been dynamically compactified to a circle of radius $R$. The flow of time changes as we move inside this circle, so that, e.g., the speed of light as measured by a four-dimensional observer may vary if the photon is also moving along the fifth dimension.

A quick glance at the metric~(\ref{eq:FW}) shows that there are two singular points, $w = \pm \pi R / 2$.
As the discussion in section~\ref{sec:prod} will show, these are bound to be coordinate singularities.
Indeed, a photon moving only along the circular dimension has a velocity
\begin{equation}
 \frac{dw}{dt} = \cos \left( \frac{w}{R} \right),
\end{equation}
so that it always moves away from $w = - \pi R / 2$ and towards $w = + \pi R / 2$.
As measured by an external observer, however, it takes an infinite amount of time for the photon to reach $w = + \pi R / 2$
(proper time for the photon is always zero),
so that the singularity is avoided.

\subsection{Shrinking Fifth Dimension}
\label{sec:shrink}

A PH$-$ solution (see table~\ref{tab:alt} and figure~\ref{fig:sols}) can take on the form
\begin{equation}
ds^{2} = - dt^{2} + d\Sigma^{2} + e^{-2t/\tau} dw^{2},
\label{eq:PT}
\end{equation}
where $\tau = l \sqrt{\xi}$ [cf.~eq.~(\ref{eq:xi})] and $\Sigma$ has constant negative curvature.

The spacetime described by the line element~(\ref{eq:PT}) features an exponentially shrinking fifth dimension, which after some time leaves an effectively four-dimensional spacetime. The time scale for this shrinking is set by $\tau$.

An arbitrarily slow shrinking ($\tau \rightarrow \infty$) can be accomplished only be coping with a highly curved ($K \rightarrow \infty$) three-manifold $\Sigma$. This case corresponds to $\chi \rightarrow 0$, i.e., the least similar one to GR.

\subsection{The AdS/CS limit}
\label{sec:cslimit}

The solutions discussed in sections~\ref{sec:circ} and~\ref{sec:shrink} have both remarkable properties when the theory is close to the AdS/CS limit $\chi = -1$ (see figure~\ref{fig:phase}).

For definiteness, let $\chi = -1 + \varepsilon$, with $0 < \varepsilon \ll 1$.
We then have $\xi = \varepsilon$ [cf.~eq.~(\ref{eq:xi})], and this has some interesting consequences.
On one hand, the radius of the circular fifth dimension from section~\ref{sec:circ} can be made arbitrarily small, $R = l \sqrt{\varepsilon} \ll l$.
On the other hand, the shrinking fifth dimension from section~\ref{sec:shrink} now shrinks extremely fast, with a characteristic time scale given by $\tau = l \sqrt{\varepsilon} \ll l$.

For both solutions, the curvature of the three-manifold $\Sigma$ is given by
$K = - \left( 1 + \varepsilon \right) / l^{2}$.

This means that even a slight deviation from an exact CS theory produces a complete breakdown of the maximally symmetric vacuum state (five-dimensional AdS spacetime) into a state where the fifth dimension either shrinks extremely fast or is compactified into an arbitrarily small circle, while ordinary three-space has roughly the same constant negative curvature as in the CS case.

\section{Maximal Extension and Product Spaces}
\label{sec:prod}

\subsection{Maximal Extension}
\label{sec:maxext}

In section~\ref{sec:sols} we showed that EGB gravity admits vacuum solutions of the form
[cf.~eq.~(\ref{eq:simpansatz})]
\begin{equation}
ds^{2} =
- f^{2} \left( w \right) dt^{2}
+ d\Sigma^{2}
+ p^{2} \left( t \right) dw^{2},
\label{eq:simpansatz2}
\end{equation}
where $d \Sigma$ is the line element of a constant-curvature three-manifold $\Sigma$ and the possible $f$ and $p$ functions are listed in table~\ref{tab:alt} for $\chi^{2} < 1$ and in table~\ref{tab:alt2} for $\chi^{2} > 1$.

Let $\Omega$ be the manifold spanned by $\left( t, w \right)$.
A direct calculation shows that $\Omega$ has \emph{constant curvature} $L$ given by
\begin{equation}
 L = \frac{1}{\xi l^{2}},
\end{equation}
where $\xi$ is defined in eq.~(\ref{eq:xi}).
The $\left( t, w \right)$ coordinate system, however, covers only part of this constant-curvature two-dimensional spacetime.
Below we show this explicitly for the two examples given in section~\ref{sec:examples}.

\begin{figure*}[tbhp]
 \centering
 \includegraphics[width=.3\textwidth]{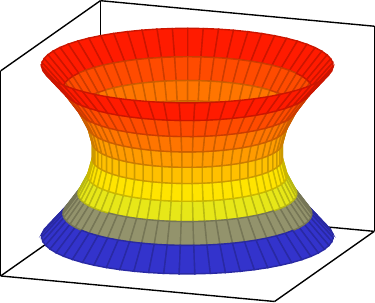}
 \hfill
 \includegraphics[width=.3\textwidth]{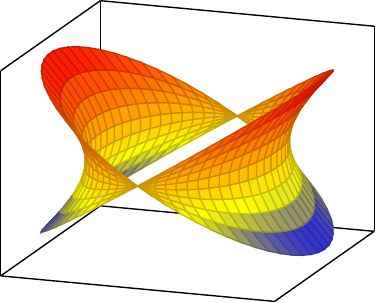}
 \hfill
 \includegraphics[width=.3\textwidth]{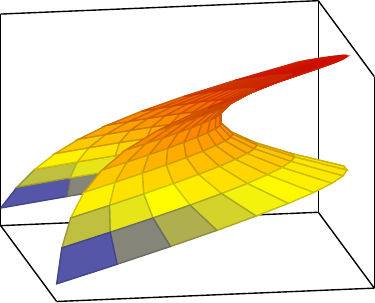}
 \caption{Left: Two-dimensional de~Sitter spacetime can be visualized as the hyperboloid $x^{2} + y^{2} - z^{2} = \xi l^{2}$, with $\xi > 0$, embedded in three-dimensional Minkowski spacetime, with the metric $ds^{2} = dx^{2} + dy^{2} - dz^{2}$. Center: the ``circular fifth dimension'' spacetime of section~\ref{sec:circ} corresponds to the region $\left| z/x \right| < 1$ of the hyperboloid. Right: the ``shrinking fifth dimension'' spacetime of section~\ref{sec:shrink} corresponds to the region $y-z>0$ of the same hyperboloid.}
 \label{fig:hbs}
\end{figure*}

Take the ``circular fifth dimension'' solution of section~\ref{sec:circ} and set
\begin{align}
 x & = R \cosh \left( \frac{t}{R} \right) \cos \left( \frac{w}{R} \right), \label{eq:x1} \\
 y & = R \sin \left( \frac{w}{R} \right), \label{eq:y1} \\
 z & = R \sinh \left( \frac{t}{R} \right) \cos \left( \frac{w}{R} \right). \label{eq:z1}
\end{align}
Eqs.~(\ref{eq:x1})--(\ref{eq:z1}) define an \emph{embedding} of the two-dimen\-sional manifold $\Omega$ in three-dimensional Minkowski spacetime,
with the metric $ds^{2} = dx^{2} + dy^{2} - dz^{2}$.
The embedding satisfies
\begin{equation}
 x^{2} + y^{2} - z^{2} = R^{2},
 \label{eq:hb}
\end{equation}
but not the whole hyperboloid~(\ref{eq:hb}) is probed by $\left( t, w \right)$.
Indeed, eq.~(\ref{eq:y1}) imposes $\left| y \right| \leq R$
[which is consistent with the condition $\left| z/x \right| < 1$ that comes from taking the quotient of eqs.~(\ref{eq:z1}) and~(\ref{eq:x1})] as a restriction on $\left( x,y,z \right)$.
This means that the ``circular fifth dimension'' solution of section~\ref{sec:circ} can be visualized as the $\left| y \right| \leq R$ region of the hyperboloid~(\ref{eq:hb}) (see figure~\ref{fig:hbs}).

An analogous phenomenon occurs for the ``shrinking fifth dimension'' solution from section~\ref{sec:shrink}.
Here the embedding equations read
\begin{align}
 x & = w e^{-t/\tau}, \label{eq:x2} \\
 y & = \frac{\tau}{2} \left[
       e^{t/\tau} - \left( \frac{w^{2}}{\tau^{2}} - 1 \right) e^{-t/\tau} \right], \label{eq:y2} \\
 z & = \frac{\tau}{2} \left[
       e^{t/\tau} - \left( \frac{w^{2}}{\tau^{2}} + 1 \right) e^{-t/\tau} \right], \label{eq:z2}
\end{align}
and a straightforward calculation shows that they too satisfy
$x^{2} + y^{2} - z^{2} = \tau^{2}$ (recall that $R=\tau$).
Subtracting eqs.~(\ref{eq:y2}) and~(\ref{eq:z2}) we get the constraint $y-z>0$.
This means that the ``shrinking fifth dimension'' solution of section~\ref{sec:shrink} can be visualized as the $y-z>0$ region of the hyperboloid~(\ref{eq:hb}) (see figure~\ref{fig:hbs}).

\begin{figure}
 \centering
 \includegraphics[width=.9\columnwidth]{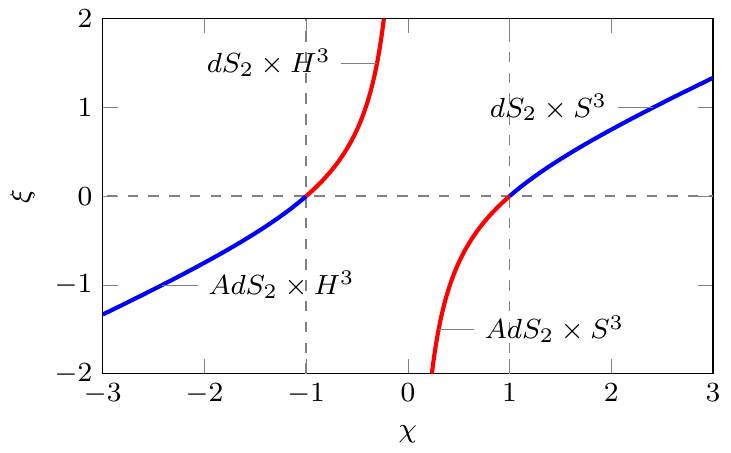}
 \caption{When maximally extended, the solutions listed in table~\ref{tab:alt} and table~\ref{tab:alt2} all have the form $\Omega \times \Sigma$, where $\Omega$ is a two-dimensional manifold of constant curvature $L=1/\xi l^{2}$ and $\Sigma$ is a three-dimensional manifold of constant curvature $K=1/\chi l^{2}$. Here we have assumed that $\Omega$ and $\Sigma$ possess, respectively, Lorentzian and Riemannian signature, but the opposite case is also possible.}
 \label{fig:solsmax}
\end{figure}

The above considerations make it clear that the maximal extension for both cases (and indeed, for all solutions with $\xi > 0$)
is the full hyperboloid $x^{2} + y^{2} - z^{2} = \xi l^{2}$, i.e., two-dimensional de~Sitter spacetime.
The constrained hyperboloids corresponding to the solutions of section~\ref{sec:circ} and section~\ref{sec:shrink} are shown in figure~\ref{fig:hbs}.
A similar reasoning applies to all solutions listed in table~\ref{tab:alt} and table~\ref{tab:alt2}.
This means that all our solutions can be extended to a product of the form $\Omega \times \Sigma$, where both manifolds have constant curvature.
Figure~\ref{fig:solsmax} shows the maximal extension catalog.

Solutions of the form $M_{2} \times M_{n-2}$, where both manifolds are maximally symmetric, are known as Nariai-type solutions, and have been independently studied in the context of Lovelock gravity by Maeda, Ray and Willison~\cite{Mae11a}.

As a historical note, the half hyperboloid obtained by enforcing the $y-z>0$ constraint is essentially a two-dimensional version of the spacetime of the steady-state model of the universe proposed by Bondi, Gold and Hoyle~\cite{Bon48,Hoy48}, with time running in the opposite direction.

\subsection{Product Spaces}
\label{sec:prospa}

Even though we started our search with a fairly complicated warped-product ansatz [see eq.~(\ref{eq:ansatz})], all solutions we found actually display a simple product structure $\Omega \times \Sigma$, where $\Omega$ and $\Sigma$ are constant-curvature manifolds of dimension two and three, respectively (see section~\ref{sec:maxext}).
A question naturally emerges: are there any other products of constant-curvature manifolds that arise as solutions of EGB gravity with $\chi^{2} < 1$?
The answer is in the negative, and here we explain why.
Before going into the details, the fact that only the ``$3+2$'' factorization, among all the possibilities, is relevant for EGB gravity, gives further support to the idea that these solutions indeed correspond to the vacuum state of the theory with $\chi^{2} < 1$.

All possible products of constant-curvature manifolds that yield a five-dimensional spacetime are in a one-to-one correspondence with the integer partitions of 5, which number 7 in total. The ``elementary'' partition $5=5$ corresponds to the case of a five-dimensional constant-curvature manifold, which only exists in regions~I and~II of figure~\ref{fig:phase}.

The first possibility is a $4+1$ factorization, where the four-dimensional manifold has constant curvature $K$.
The EGB field equations are solved by $K=1/\chi l^{2}$, but only for two values of $\chi$, namely, $\chi = \pm 1$, i.e., the CS case.%

A second interesting possibility is the $5=2+2+1$ partition.
Here a solution exists only when $\chi = \pm \sqrt{3}$, which pushes us into region~II in figure~\ref{fig:phase}.
The two two-dimensional manifolds must have the same constant curvature
$\pm \sqrt{3}/l^{2}$.

The three remaining partitions,
$5=3+1+1$,
$5=2+1+1+1$, and
$5=1+1+1+1+1$
all yield trivial results: the only allowed solution is Minkowski spacetime
when the cosmological constant vanishes (i.e., along the vertical line separating regions~I and~II in figure~\ref{fig:phase}).

This brief but thorough analysis shows that the only product space allowed in EGB gravity with $\chi^{2} < 1$ is $5=3+2$, which lends credence to the idea that this solution indeed corresponds to the vacuum state.

\subsection{No warped $4+1$ product when $\chi^{2} < 1$}

As explained in section~\ref{sec:prospa}, a simple $4+1$ product is a solution of the EGB field equations only in the CS case, i.e., when $\chi = \pm 1$.
One may ask whether a \emph{warped} $4+1$ ansatz would allow us to find a new vacuum candidate when $\chi^{2} < 1$.
Indeed, the ansatz
\begin{equation}
ds^{2} = f^{2} \left( w \right) d \Sigma^{2} + dw^{2}
\end{equation}
provides two different (but related) solutions: either
$f \left( w \right) = \sin \left( w/R \right)$, or
$f \left( w \right) = \sinh \left( w/R \right)$.
Here $R$ is implicitly defined by the equation
\begin{equation}
\chi = \pm \frac{1}{2} \left[ \left( \frac{R}{l} \right)^{2} + \left( \frac{l}{R} \right)^{2} \right],
\end{equation}
where the plus sign applies to the circular solution and the minus sign to the hyperbolic one.
In both cases we have $\chi^{2} \ge 1$, which lies outside region~III in figure~\ref{fig:phase}.
This means that there exist no warped $4+1$ solutions when $\chi^{2} < 1$, i.e.,
in the region of parameter space where a constant-curvature five-dimensional spacetime is not possible.

\section{Conclusions}
\label{sec:final}

In this paper we have analyzed possible candidates for the vacuum state of the EGB theory in five dimensions, in the particular case where the coupling constants of the theory do not allow for a constant-curvature solution.
Five-dimensional spacetime has a dual appeal; on one hand, it is the lowest dimension where the EGB theory can be defined, and on the other hand, it is interesting in view of the AdS/CFT correspondence~\cite{Aha99}.

The general picture we find is that of a product of a three-manifold $\Sigma$ and a two-manifold $\Omega$, both with constant curvatures
\begin{align}
 K & = \frac{1}{\chi l^{2}}, \\
 L & = \frac{1}{\xi l^{2}},
\end{align}
respectively. Here $l$ is a global length scale, $\chi$ is a dimensionless parameter that measures the relative strength of the EH term in the EGB action, and $\xi$ is defined by [cf.~eq.~(\ref{eq:xi})]
\begin{equation}
 \xi = \frac{1}{2} \left( \chi - \frac{1}{\chi} \right).
\end{equation}
No other simple product of constant-curvature manifolds is allowed
as a solution of the EGB field equations in the $\chi^{2} < 1$ regime.
The $3+2$ factorization reported here is unique within this class of solutions.
A warped $4+1$ product exists only when $\chi^{2} \ge 1$.

Two solutions have seemed particularly interesting to us, since both can be regarded as effectively four-dimen\-sional, with the fifth dimension either being compactified to a circle or shrinking exponentially with time.
These correspond to particular truncations of the product manifold $\Omega \times \Sigma$.

Furthermore, we have found that even a slight deviation from the exact AdS/CS theory (with $\chi = -1$) produces a complete breakdown of the maximally symmetric vacuum state (five-dimensional AdS spacetime) into one of the above-mentioned solutions.
Specifically, we find that, for $\chi = -1 + \varepsilon$, with $0 < \varepsilon \ll 1$,
either the circle has an arbitrarily small radius $R = l \sqrt{\varepsilon} \ll l$ or the fifth dimension shrinks extremely fast,
with a characteristic time scale $\tau = l \sqrt{\varepsilon} \ll l$.
In both cases the three-manifold $\Sigma$ has constant negative curvature only slightly different from five-dimensional AdS spacetime, namely $K = - \left( 1 + \varepsilon \right) / l^{2}$.

It is somehow puzzling, or perhaps revealing, that both these effectively four-dimen\-sional solutions occur when $\chi$ is close to the AdS/CS value of $\chi = -1$.
There seem to be no similarly interesting (i.e., that display a dynamical dimensional reduction to four dimensions) solutions with $\chi$ close to the dS/CS value of $\chi = +1$.
It must be stressed that this is completely unexpected from the point of view of CS dynamics, where the vacuum state is five-dimensional (A)dS spacetime.

There are several foreseeable extensions for this work.
In trying to answer the question why our universe has four observable dimensions, the first thing one must do is stop \emph{assuming} it has four from the outset.
Studying theories in five dimensions is clearly a step forward, but far from being enough.
Ideally, one would like the theory itself to select the dimension of spacetime, much like it occurs in String theory, for instance.
Unfortunately, String theory does not provide a natural explanation of why precisely six of the ten dimensions of spacetime should be compactified.
In this spirit, it would be interesting to see if effectively four-dimensional spacetimes emerge as the vacuum state of EGB gravity in dimensions higher than five.
While the field equations are essentially the same as in $d=5$, the ansatz~(\ref{eq:ansatz}) clearly must be modified.
As a first step in this direction, a full analysis of products of constant-curvature manifolds as solutions of EGB gravity in dimensions higher than five is currently underway and will be reported elsewhere.
On the other hand, third and higher-order Lovelock terms become available as we increase the spacetime dimension.
In $d=11$, for instance, the most general Lovelock Lagrangian includes up to fifth-order powers of the curvature.
Finding the regime where no constant-curvature solutions exist in this eleven-dimensional case amounts to solving a fifth-order polynomial.
This will lead to a much richer set of possibilities than the one we have discussed here.
We expect the techniques developed by Camanho and Edelstein~\cite{Cam11} to be relevant here.

Our solutions are simple enough to hold them as strong candidates for the vacuum state of the EGB theory with $\chi^{2} < 1$,
but there may be others.%
\footnote{Homogeneous spaces~\cite{Ste03,Gri09} appear to be promising candidates.}
The main issue that needs addressing is the question of stability.
Being a quad\-ratic theory, the definition of energy in EGB gravity is different from that of GR, and specific methods to deal with it have been developed in the literature~\cite{Des89,Mor04a,Mor04b}.
In particular, the formalism of Ref.~\cite{Aro99b} may be useful for EGB gravity in six dimensions, while higher-order Lovelock terms would be required for its application in eight or higher dimensions.
Even in the ``usual'' case where two constant-curvature solutions exist, deciding which one of them can be regarded as the true vacuum is nontrivial~\cite{Bou85,Des89,Des02a,Des02b,Des07,Cha08}.
We would like to stress that, while most surely the techniques developed in the literature will be useful in deciding whether any of our solutions can be regarded as the vacuum, the conclusions that have been reached so far are inapplicable in our case, since they refer to constant-curvature spaces that are not allowed in the $\chi^{2} < 1$ regime.
A more direct approach to the stability question would be to explicitly compute the propagation of linear perturbations around these vacua.
Here one must take care to consistently define what can be regarded as scalar, vector and tensor perturbations, in a manner analogous as to what is done in Cosmology when perturbing the FRW metric (see, e.g., Ref.~\cite{Muk92}).
These issues are currently under investigation.

\begin{acknowledgments}
Enlightening conversations with
M.~Ba\~{n}ados,
J.~A.~Helay\"{e}l-Neto,
J.~Oliva and
J.~Zanelli
are gratefully acknowledged.
F.~I.\ would like to thank
J.~M.~Izquierdo for his kind hospitality at the Universidad de Valladolid, Spain.
E.~R.\ is grateful to
J.~A.~Helay\"{e}l-Neto for his warm hospitality at the Centro Brasileiro de Pesquisas F\'{\i}sicas (CBPF) in Rio de Janeiro, Brazil.
The authors were partly supported by Direcci\'{o}n de Perfeccionamiento y Postgrado, Universidad Cat\'{o}lica de la Sant\'{\i}sima Concepci\'{o}n, Concepci\'{o}n, Chile.
E.~R.\ was supported by Conselho Nacional de Desenvolvimento Científico e Tecnológico (CNPq), Brazil, through grant 190772/2011-5.
\end{acknowledgments}

\bibliographystyle{utphys}
\bibliography{biblio2013}

\providecommand{\href}[2]{#2}\begingroup\raggedright\begin{thebibliography}{10}

\bibitem{Car07}
S.~M. Carroll, ``{Lecture Notes on General Relativity},''
  \href{http://arxiv.org/abs/gr-qc/9712019}{{\ttfamily arXiv:gr-qc/9712019}}.

\bibitem{Lov71}
D.~Lovelock, ``The {Einstein} tensor and its generalizations,''
  \href{http://dx.doi.org/10.1063/1.1665613}{{\em J. Math. Phys.} {\bfseries
  12} (1971) 498}.

\bibitem{Lan38}
C.~Lanczos, ``A remarkable property of the {Riemann--Christoffel} tensor in
  four dimensions,'' {\em Ann. Math.} {\bfseries 39} (1938) 842.
  \url{http://www.jstor.org/stable/1968467}.

\bibitem{Aro99a}
R.~Aros, M.~Contreras, R.~Olea, R.~Troncoso, and J.~Zanelli, ``Conserved
  charges for gravity with locally {AdS} asymptotics,''
  \href{http://dx.doi.org/10.1103/PhysRevLett.84.1647}{{\em Phys. Rev. Lett.}
  {\bfseries 84} (2000) 1647--1650},
  \href{http://arxiv.org/abs/gr-qc/9909015}{{\ttfamily arXiv:gr-qc/9909015}}.

\bibitem{Aro99b}
R.~Aros, M.~Contreras, R.~Olea, R.~Troncoso, and J.~Zanelli, ``Conserved
  charges for even dimensional asymptotically {AdS} gravity theories,''
  \href{http://dx.doi.org/10.1103/PhysRevD.62.044002}{{\em Phys. Rev. D}
  {\bfseries 62} (2000) 044002},
  \href{http://arxiv.org/abs/hep-th/9912045}{{\ttfamily arXiv:hep-th/9912045}}.

\bibitem{Zwi85}
B.~Zwiebach, ``Curvature squared terms and string theories,''
  \href{http://dx.doi.org/10.1016/0370-2693(85)91616-8}{{\em Phys. Lett. B}
  {\bfseries 156} (1985) 315--317}.

\bibitem{Bou85}
D.~G. Boulware and S.~Deser, ``String-generated gravity models,''
  \href{http://dx.doi.org/10.1103/PhysRevLett.55.2656}{{\em Phys. Rev. Lett.}
  {\bfseries 55} (1985) 2656--2660}.

\bibitem{Zum86}
B.~Zumino, ``Gravity theories in more than four dimensions,''
  \href{http://dx.doi.org/10.1016/0370-1573(86)90076-1}{{\em Phys. Rep.}
  {\bfseries 137} (1986) 109--114}.

\bibitem{Wil86}
D.~L. Wiltshire, ``{Spherically symmetric solutions of Einstein--Maxwell theory
  with a Gauss--Bonnet term},''
\href{http://dx.doi.org/10.1016/0370-2693(86)90681-7}{{\em Phys. Lett.}
  {\bfseries B169} (1986) 36}.

\bibitem{Aha99}
O.~Aharony, S.~S. Gubser, J.~M. Maldacena, H.~Ooguri, and Y.~Oz, ``{Large $N$
  field theories, string theory and gravity},''
  \href{http://dx.doi.org/10.1016/S0370-1573(99)00083-6}{{\em Phys. Rept.}
  {\bfseries 323} (2000) 183--386},
\href{http://arxiv.org/abs/hep-th/9905111}{{\ttfamily arXiv:hep-th/9905111
  [hep-th]}}.

\bibitem{Noj06}
S.~Nojiri and S.~D. Odintsov, ``Introduction to modified gravity and
  gravitational alternative for dark energy,''
  \href{http://dx.doi.org/10.1142/S0219887807001928}{{\em Int. J. Geom. Meth.
  Mod. Phys.} {\bfseries 4} (2007) 115--146},
  \href{http://arxiv.org/abs/hep-th/0601213}{{\ttfamily arXiv:hep-th/0601213}}.

\bibitem{Mad85}
J.~Madore, ``{Kaluza--Klein theory with the Lanczos Lagrangian},''
\href{http://dx.doi.org/10.1016/0375-9601(85)90773-X}{{\em Phys. Lett.}
  {\bfseries A110} (1985) 289}.

\bibitem{Mul85}
F.~Müller-Hoissen, ``{Spontaneous compactification with quadratic and cubic
  curvature terms},''
\href{http://dx.doi.org/10.1016/0370-2693(85)90202-3}{{\em Phys. Lett.}
  {\bfseries B163} (1985) 106}.

\bibitem{Whe86a}
J.~T. Wheeler, ``Symmetric solutions to the {Gauss--Bonnet} extended {Einstein}
  equations,'' \href{http://dx.doi.org/10.1016/0550-3213(86)90268-3}{{\em Nucl.
  Phys. B} {\bfseries 268} (1986) 737--746}.

\bibitem{Whe86b}
J.~T. Wheeler, ``Symmetric solutions to the maximally {Gauss--Bonnet} extended
  {Einstein} equations,''
  \href{http://dx.doi.org/10.1016/0550-3213(86)90388-3}{{\em Nucl. Phys. B}
  {\bfseries 273} (1986) 732--748}.

\bibitem{Cve01}
M.~Cveti{\v{c}}, S.~Nojiri, and S.~D. Odintsov, ``Black hole thermodynamics and
  negative entropy in {de Sitter} and anti-{de Sitter}
  {Einstein--Gauss--Bonnet} gravity,''
  \href{http://dx.doi.org/10.1016/S0550-3213(02)00075-5}{{\em Nucl. Phys. B}
  {\bfseries 628} (2002) 295--330},
  \href{http://arxiv.org/abs/hep-th/0112045}{{\ttfamily arXiv:hep-th/0112045}}.

\bibitem{Cai01}
R.-G. Cai, ``{Gauss--Bonnet} black holes in {AdS} spaces,''
  \href{http://dx.doi.org/10.1103/PhysRevD.65.084014}{{\em Phys. Rev. D}
  {\bfseries 65} (2002) 084014},
  \href{http://arxiv.org/abs/hep-th/0109133}{{\ttfamily arXiv:hep-th/0109133}}.

\bibitem{Cho01}
Y.~M. Cho, I.~P. Neupane, and P.~S. Wesson, ``No ghost state of {Gauss--Bonnet}
  interaction in warped background,''
  \href{http://dx.doi.org/10.1016/S0550-3213(01)00579-X}{{\em Nucl. Phys. B}
  {\bfseries 621} (2002) 388--412},
\href{http://arxiv.org/abs/hep-th/0104227}{{\ttfamily arXiv:hep-th/0104227
  [hep-th]}}.

\bibitem{Cha02}
C.~Charmousis and J.-F. Dufaux, ``General {Gauss--Bonnet} brane cosmology,''
  \href{http://dx.doi.org/10.1088/0264-9381/19/18/304}{{\em Class. Quant.
  Grav.} {\bfseries 19} (2002) 4671--4682},
  \href{http://arxiv.org/abs/hep-th/0202107}{{\ttfamily arXiv:hep-th/0202107}}.

\bibitem{Can07a}
F.~Canfora, A.~Giacomini, and S.~Willison, ``Some exact solutions with torsion
  in 5-d {Einstein}--{Gauss}--{Bonnet} gravity,''
  \href{http://dx.doi.org/10.1103/PhysRevD.76.044021}{{\em Phys. Rev. D}
  {\bfseries 76} (2007) 044021},
  \href{http://arxiv.org/abs/0706.2891}{{\ttfamily arXiv:0706.2891 [gr-qc]}}.

\bibitem{Can07b}
F.~Canfora, A.~Giacomini, and R.~Troncoso, ``Black holes, parallelizable
  horizons and half-{BPS} states for the {Einstein}--{Gauss}--{Bonnet} theory
  in five dimensions,''
  \href{http://dx.doi.org/10.1103/PhysRevD.77.024002}{{\em Phys. Rev. D}
  {\bfseries 77} (2008) 024002},
  \href{http://arxiv.org/abs/0707.1056}{{\ttfamily arXiv:0707.1056 [hep-th]}}.

\bibitem{Dot10}
G.~Dotti, J.~Oliva, and R.~Troncoso, ``Static solutions with nontrivial
  boundaries for the {Einstein--Gauss--Bonnet} theory in vacuum,''
  \href{http://dx.doi.org/10.1103/PhysRevD.82.024002}{{\em Phys. Rev. D}
  {\bfseries 82} (2010) 024002},
  \href{http://arxiv.org/abs/1004.5287}{{\ttfamily arXiv:1004.5287 [hep-th]}}.

\bibitem{Mae11b}
H.~Maeda, ``{Gauss--Bonnet} braneworld redux: A novel scenario for the bouncing
  universe,'' \href{http://dx.doi.org/10.1103/PhysRevD.85.124012}{{\em Phys.
  Rev. D} {\bfseries 85} (2012) 124012},
  \href{http://arxiv.org/abs/1112.5178}{{\ttfamily arXiv:1112.5178}}.

\bibitem{Gol12}
S.~Golod and T.~Piran, ``Choptuik's critical phenomenon in
  {Einstein--Gauss--Bonnet} gravity,''
  \href{http://dx.doi.org/10.1103/PhysRevD.85.104015}{{\em Phys. Rev. D}
  {\bfseries 85} (2012) 104015},
  \href{http://arxiv.org/abs/1201.6384}{{\ttfamily arXiv:1201.6384 [gr-qc]}}.

\bibitem{Bri12}
Y.~Brihaye and B.~Hartmann, ``Hairy charged {Gauss--Bonnet} solitons and black
  holes,'' \href{http://dx.doi.org/10.1103/PhysRevD.85.124024}{{\em Phys. Rev.
  D} {\bfseries 85} (2012) 124024},
  \href{http://arxiv.org/abs/1203.3109}{{\ttfamily arXiv:1203.3109 [gr-qc]}}.

\bibitem{Des89}
S.~Deser and Z.~Yang, ``Energy and stability in {Einstein--Gauss--Bonnet}
  models,'' \href{http://dx.doi.org/10.1088/0264-9381/6/5/001}{{\em Class.
  Quant. Grav.} {\bfseries 6} (1989) L83}.

\bibitem{Des02a}
S.~Deser and B.~Tekin, ``Gravitational energy in quadratic curvature
  gravities,'' \href{http://dx.doi.org/10.1103/PhysRevLett.89.101101}{{\em
  Phys. Rev. Lett.} {\bfseries 89} (2002) 101101},
  \href{http://arxiv.org/abs/hep-th/0205318}{{\ttfamily arXiv:hep-th/0205318}}.

\bibitem{Des02b}
S.~Deser and B.~Tekin, ``Energy in generic higher curvature gravity theories,''
  \href{http://dx.doi.org/10.1103/PhysRevD.67.084009}{{\em Phys. Rev. D}
  {\bfseries 67} (2003) 084009},
  \href{http://arxiv.org/abs/hep-th/0212292}{{\ttfamily arXiv:hep-th/0212292}}.

\bibitem{Des07}
S.~Deser and B.~Tekin, ``New energy definition for higher curvature
  gravities,'' \href{http://dx.doi.org/10.1103/PhysRevD.75.084032}{{\em Phys.
  Rev. D} {\bfseries 75} (2007) 084032},
  \href{http://arxiv.org/abs/gr-qc/0701140}{{\ttfamily arXiv:gr-qc/0701140}}.

\bibitem{Cha08}
C.~Charmousis and A.~Padilla, ``The instability of vacua in {Gauss--Bonnet}
  gravity,'' \href{http://dx.doi.org/10.1088/1126-6708/2008/12/038}{{\em JHEP}
  {\bfseries 0812} (2008) 038},
  \href{http://arxiv.org/abs/0807.2864}{{\ttfamily arXiv:0807.2864 [hep-th]}}.

\bibitem{Cha89}
A.~H. Chamseddine, ``Topological gauge theory of gravity in five and all odd
  dimensions,'' \href{http://dx.doi.org/10.1016/0370-2693(89)91312-9}{{\em
  Phys. Lett. B} {\bfseries 233} (1989) 291--294}.

\bibitem{Mae11a}
H.~Maeda, S.~Willison, and S.~Ray, ``Lovelock black holes with maximally
  symmetric horizons,''
  \href{http://dx.doi.org/10.1088/0264-9381/28/16/165005}{{\em Class. Quant.
  Grav.} {\bfseries 28} (2011) 165005},
  \href{http://arxiv.org/abs/1103.4184}{{\ttfamily arXiv:1103.4184}}.

\bibitem{Bon48}
H.~Bondi and T.~Gold, ``The steady-state theory of the expanding universe,''
  {\em Mon. Not. Roy. Ast. Soc.} {\bfseries 108} (1948) 252--270.
  \url{http://adsabs.harvard.edu/abs/1948MNRAS.108..252B}.

\bibitem{Hoy48}
F.~Hoyle, ``A new model for the expanding universe,'' {\em Mon. Not. Roy. Ast.
  Soc.} {\bfseries 108} (1948) 372--382.
  \url{http://adsabs.harvard.edu/abs/1948MNRAS.108..372H}.

\bibitem{Cam11}
X.~O. Camanho and J.~D. Edelstein, ``A {Lovelock} black hole bestiary,''
  \href{http://dx.doi.org/10.1088/0264-9381/30/3/035009}{{\em Class. Quant.
  Grav.} {\bfseries 30} (2013) 035009},
\href{http://arxiv.org/abs/1103.3669}{{\ttfamily arXiv:1103.3669 [hep-th]}}.

\bibitem{Ste03}
H.~Stephani, D.~Kramer, M.~MacCallum, C.~Hoenselaers, and E.~Herlt, {\em {Exact
  Solutions of Einstein's Field Equations}}.
\newblock Cambridge University Press, Cambridge, UK, 2nd~ed., 2003.

\bibitem{Gri09}
J.~B. Griffiths and J.~Podolsk{\'{y}}, {\em {Exact Space-Times in Einstein's
  General Relativity}}.
\newblock Cambridge University Press, Cambridge, UK, 1st~ed., 2009.

\bibitem{Mor04a}
P.~Mora, R.~Olea, R.~Troncoso, and J.~Zanelli, ``Finite action principle for
  {Chern--Simons} {AdS} gravity,''
  \href{http://dx.doi.org/10.1088/1126-6708/2004/06/036}{{\em JHEP} {\bfseries
  0406} (2004) 036}, \href{http://arxiv.org/abs/hep-th/0405267}{{\ttfamily
  arXiv:hep-th/0405267}}.

\bibitem{Mor04b}
P.~Mora, R.~Olea, R.~Troncoso, and J.~Zanelli, ``Vacuum energy in
  odd-dimensional {AdS} gravity,''
  \href{http://arxiv.org/abs/hep-th/0412046}{{\ttfamily arXiv:hep-th/0412046}}.

\bibitem{Muk92}
V.~F. Mukhanov, H.~A. Feldman, and R.~H. Brandenberger, ``Theory of
  cosmological perturbations,''
  \href{http://dx.doi.org/10.1016/0370-1573(92)90044-Z}{{\em Phys. Rept.}
  {\bfseries 215} (1992) 203--333}.

\end{thebibliography}\endgroup

\end{document}